\documentclass[twocolumn,showpacs,pre]{revtex4}
\usepackage{graphicx}

\begin{document}

\title{Radiation pattern of a classical dipole in a photonic crystal: photon
focusing}

\author{Dmitry N. Chigrin}
\email{chigrin@uni-wuppertal.de}
\affiliation{Institute of Materials Science and Department of Electrical and Information
Engineering, University of Wuppertal, Gauss-str.~20, 42097 Wuppertal,
Germany}

\begin{abstract}
The asymptotic analysis of the radiation pattern of a classical dipole
in a photonic crystal possessing an incomplete photonic bandgap is
presented. The far-field radiation pattern demonstrates a strong modification
with respect to the dipole radiation pattern in vacuum. Radiated power
is suppressed in the direction of the spatial stopband and strongly
enhanced in the direction of the group velocity, which is stationary
with respect to a small variation of the wave vector. An effect of
radiated power enhancement is explained in terms of \emph{photon focusing}.
Numerical example is given for a square-lattice two-dimensional photonic
crystal. Predictions of asymptotic analysis are substantiated with
finite-difference time-domain calculations, revealing a reasonable
agreement.
\end{abstract}

\pacs{42.70.Qs; 42.25.Fx; 42.50.Pq; 81.05.Zx}

\maketitle

\section{Introduction}

Purcell~\cite{purcell46} was the first who pointed out, that the
spontaneous emission of an atom or a molecule depends on its environment.
Since then, an influence of non-trivial boundary conditions in the
vicinity of an excited atom on its emissive properties has been the
subject of active research \cite{bookYokayama95,bookBW95,bookDowling97}.
Important examples of such an influence are an enhancement and inhibition
of the spontaneous emission by a resonant environment~\cite{purcell46},
e.g., microcavity. These phenomena were first demonstrated by Goy
\emph{et}~\emph{al.}~\cite{goy83} and Kleppner~\cite{kleppner81},
respectively, and continue to be the subject of intense research not
only due to their contribution to the better understanding of the
light matter interaction, but, to a great extend, due to the practical
importance of controlling the light emission process. Light-emitting
diodes \cite{benisty98a,benisty98b,delbeke02} and thresholdless lasers
\cite{yokoyama89,bjork93,bjork94} are just a few examples, where
the light extraction and the spontaneous emission control by mean
of optical microcavity leads to improved performance.

Dielectric periodic medium, also called photonic crystal \cite{bookJDJ,bookSakoda},
is a good example of non-trivial boundary conditions on electromagnetic
field. Such an inhomogeneous medium can possess a complete photonic
bandgap, i.e., a continuous spectral range within which linear propagation
of light is prohibited in all spatial directions. One of the consequence
is an inhibited spontaneous emission for the atomic transition frequency
inside the complete photonic bandgap \cite{Byk72,eli87,john87}. There
are no electromagnetic modes avaliable to carry the energy away from
the atom at complete photonic bandgap frequencies. Although an existence
of complete photonic bandgap usually requires a high index ($n>3$)
dielectric materials arranged in a three-dimensional (3D) lattice
\cite{bookJDJ,bookSakoda}, photonic crystals are proven to be useful
artificial materials to modify the light emission even in the absence
of complete photonic bandgap. For example, it was demonstrated that
the external quantum efficiency of light-emitting diodes can be significantly
improved by introducing a two-dimensional (2D) photonic crystal \cite{hirayama96,boroditsky99}.
Another example is a highly directive light source employing a 3D
photonic crystal \cite{temelkuran00,enoch96}.

An intrinsic property of photonic crystals is their complicated photonic
band structure, which can be engineered by choosing an appropriate
combination of materials and lattice geometry \cite{bookJDJ,bookSakoda}.
Being able to modify in purpose the emission rate within a specific
spectral range and simultaneously in specific directions could add
a significant flexibility in improving light sources.

A number of papers were devoted to the study of the spontaneous emission
in photonic crystals, considering emission modification using both
classical \cite{dowling92,suzuki95,sakoda96,xu00,lousse01,hermann02}
and quantum \cite{Byk72,john90,zhu00,xu00,li00,busch00,li01,woldeyohannes03}
formalisms. But, to the author knowledge, questions like modification
of the emission rate in a specific direction and modification of the
emission pattern due to the photonic crystal environment have not
been yet addressed. Special opportunities in controlling directionality
of emission exist within spectral ranges of allowed photonic bands,
where photonic crystals display strong dispersion and anisotropy.
The consequence of anisotropy is the beam steering effect \cite{russell86a,zengerle87},
which in the essence means, that the group velocity direction of the
medium's eigenmode does not necessarily coincide with its wave vector
direction. A beam steering effect known to be a reason for the number
of anomalies in an electromagnetic beam propagation inside a photonic
crystal, which are usually referred to as superprism or ultrarefractive
phenomena \cite{russell86a,zengerle87,kosaka98}. For example, an
extraordinary large or negative beam bending \cite{kosaka98}, a beam
self-collimation \cite{kosaka99,chigrin2003-1} and the photon focusing
\cite{etchegoin96,chigrin2001} were reported. The last phenomenon
is similar to the phonon focusing, phenomenon observed in the ballistic
transport of phonons in crystalline solid \cite{bookWolfe98}.

The term phonon focusing refers to the strong anisotropy of heat flux
in crystalline solid. First observed in 1969 by Taylor \emph{et}~\emph{al.}~\cite{taylor69},
phonon focusing is a property of all crystals at low temperatures.
The term {}``focusing'' does not imply a bending of particle paths,
as in the geometrical-optics sense of the term. The physical reason
of the phonon focusing is the beam steering. In particular, waves
with quite different wave vectors can have nearly the same group velocity,
so the energy flux associated with those waves bunches along certain
crystalline directions. In some special cases, a heat flux can display
intricate focusing caustics, along which flux tends to infinity \cite{bookWolfe98}.
This happens when the direction of the group velocity is stationary
with respect to a small variation of the wave vector.

One can expect, that a similar phenomenon takes place in photonic
crystals \cite{etchegoin96,chigrin2001}. An optical cousin of the
acoustic phenomenon opens a unique opportunity to design a caustics
pattern on purpose, enhancing and suppressing emission in specific
directions.

In this paper a description of an angular distribution of radiated
power of a classical dipole embedded in a photonic crystal is presented.
It is assumed that only propagating modes of the photonic crystal
contribute to the far-field radiation. The emission process is treated
using an entirely classical model, similar to one in \cite{dowling92,sakoda96}.
It is commonly accepted that a classical description leads to the
same results as an entirely quantum electrodynamical approach \cite{dowling92,xu00}.
In the classical description, the modification of spontaneous emission
is due to the radiation reaction of the back-reflected field on the
classical dipole \cite{sipe81,ford84,wylie84}. Then within the framework
of the Weisskopf-Wigner approximation \cite{glauber91,busch00}, the
spontaneous emission rate, $\Gamma$, is related to the classical
radiated power $P(\mathbf{r}_{0})=\left(\omega/2\right)\mathrm{Im[}\mathbf{d}^{*}\cdot\mathbf{E}\mathrm{(}\mathbf{r}_{0}\mathrm{)]}$
via $\Gamma=P/\hbar\omega$ \cite{ford84}, where $\mathbf{d}$ is
a real dipole moment, $\mathbf{E}(\mathbf{r}_{0})$ is a field in
the system and $\mathbf{r}_{0}$ is the dipole location. A well known
interpretation of the emission rate modification, as the dipole interaction
with the out-of-phase part of the radiation reaction field, follows
from that relation \cite{sipe81,wylie84}. In the classical picture,
a non-relativistic Lamb shift is due to the dipole interacting with
the in-phase part of the reaction field \cite{wylie84,dowling98}
and it is also seen to be a purely classical effect \cite{wylie84,dowling98}.
Although, the magnitude of the anomalous Lamb shift in a realistic
photonic crystal is actively and controversially discussed \cite{john90,zhu00,li00,li01},
this question is out of the scope of this work.

A general expressions for the field and emission
rate of the point dipole radiating in an arbitrary periodic medium
are reviewed in sections \ref{sec:dipole_field} and \ref{sec:emission_rate},
respectively. The evaluation of the asymptotic form of the radiated
field is given in section \ref{sec:asymptotic_far_field}. In section
\ref{sec:angular_distribution_of_power}, an angular distribution
of radiated power is introduced. A modification of radiation pattern
is discussed in terms of photon focusing in section \ref{sec:photon_focusing}.
A numerical example of an angular distribution of emission power radiated
from the point isotropic light source is presented in section \ref{sec:example_2D_PhC}
for the case of a two-dimensional square lattice photonic crystal
of dielectric rods in air. Summary is given in section \ref{sec:dipole:summary}.

\section{\label{sec:dipole_field}Normal mode expansion of dipole field}

In this paper, a general linear, non-magnetic, dielectric medium with
arbitrary 3D periodic dielectric function, $\varepsilon(\mathbf{r})=\varepsilon(\mathbf{r}+\mathbf{R})$,
is studied. Here $\mathbf{R}$ is a vector of the direct Bravais lattice,
$\mathbf{R}=\sum_{i}l_{i}\mathbf{a}_{i}$, $l_{i}$ is an integer
and $\mathbf{a}_{i}$ is a basis vector of the periodic lattice. It
is assumed that a medium is infinitely extended in space and that
no absorption happens. In general, presented approach is valid for
any inhomogeneous, non-absorbing medium, for which a dispersion relations
can be found in the form, $\omega=\omega(\mathbf{k})$, numerically
or analytically. Here $\mathbf{k}$ is the wave vector.

In Gaussian units, Maxwell's equations in such a medium have a form\begin{eqnarray}
 &  & \nabla\times\mathbf{E}=-\frac{1}{c}\frac{\partial\mathbf{H}}{\partial t},\label{math:Maxwell1}\\
 &  & \nabla\times\mathbf{H}=\frac{1}{c}\varepsilon(\mathbf{r})\frac{\partial\mathbf{E}}{\partial t}+\frac{4\pi}{c}\mathbf{J},\label{math:Maxwell2}\\
 &  & \nabla\cdot[\varepsilon(\mathbf{r})\mathbf{E}]=0,\label{math:Maxwell3}\\
 &  & \nabla\cdot\mathbf{H}=0.\label{math:Maxwell4}\end{eqnarray}
 Here, the electric (magnetic) field is denoted by \textbf{$\mathbf{E}$}
($\mathbf{H}$), $c$ is a speed of light in vacuum. An electromagnetic
field is produced by a current source $\mathbf{J}$ and the charge
density is zero, $\rho\equiv0$. Then one can choose the transverse
(Coulomb) gauge for the vector potential $\mathbf{A}$ in the form
\cite{glauber91}:\begin{equation}
\nabla\cdot[\varepsilon(\mathbf{r})\mathbf{A}]=0.\label{math:gauge}\end{equation}
The absence of the charge density implies that the scalar potential
$\varphi$ is zero. The electric and magnetic fields can be written
in terms of the vector potential $\mathbf{A}$ via:\begin{eqnarray}
 &  & \mathbf{E}=-\frac{1}{c}\frac{\partial\mathbf{A}}{\partial t},\label{math:EviaA}\\
 &  & \mathbf{H}=\nabla\times\mathbf{A}.\label{math:HviaA}\end{eqnarray}
Combining equations (\ref{math:EviaA}-\ref{math:HviaA}) with Maxwell's
equations (\ref{math:Maxwell1}-\ref{math:Maxwell4}) one obtains the
wave equation for the vector potential $\mathbf{A}$: \begin{equation}
\nabla\times\nabla\times\mathbf{A}+\frac{1}{c^{2}}\varepsilon(\mathbf{r})\frac{\partial^{2}\mathbf{A}}{\partial t^{2}}=\frac{4\pi}{c}\mathbf{J}.\label{math:waveEQ-A}\end{equation}
In what follows, a simplest form of the current density $\mathbf{J}$
is taken:\begin{equation}
\mathbf{J}(\mathbf{r},t)=-i\omega_{0}\mathbf{d}\delta(\mathbf{r}-\mathbf{r}_{0})e^{-i\omega_{0}t}\label{math:sourceTerm}\end{equation}
for a harmonically oscillating dipole with a frequency $\omega_{0}$
and a real dipole moment $\mathbf{d}$, located at the position $\mathbf{r}_{0}$
inside a photonic crystal, switched on at $t=0$.

The field of an arbitrary light source embedded in a periodic medium
can be constructed by a suitable superposition of the medium's eigenwaves:

\begin{equation}
\mathbf{A}(\mathbf{r},t)=\sum_{n}\int_{BZ}d^{3}\mathbf{k}_{n}C_{n\mathbf{k}}(t)\mathbf{A}_{n\mathbf{k}}(\mathbf{r}).\label{math:generalField}\end{equation}
Here $\mathbf{A}_{n\mathbf{k}}(\mathbf{r})$ and $C_{n\mathbf{k}}(t)$
are the Bloch eigenvector and the time-dependent amplitude coefficient
of the eigenwave $(n,\mathbf{k})$, respectively. The form of the
amplitude coefficient is defined by the particular nature of the light
source. The integration is performed over the first Brillouin zone
(BZ) of the crystal and the summation is carried out over different
photonic bands, where $n$ is the band index.

Eigenwaves $\mathbf{A}_{n\mathbf{k}}(\mathbf{r})$ satisfy the homogeneous
wave equation\begin{equation}
\nabla\times\nabla\times\mathbf{A}_{n\mathbf{k}}-\frac{\omega_{n\mathbf{k}}^{2}}{c^{2}}\varepsilon(\mathbf{r})\mathbf{A}_{n\mathbf{k}}=0\label{math:waveEQ-Ank}\end{equation}
and \textbf{}also fulfill the orthogonalization, normalization and
closure conditions given by:\begin{eqnarray}
 &  & \int_{V}d^{3}\mathbf{r}\varepsilon(\mathbf{r})\mathbf{A}_{n\mathbf{k}}(\mathbf{r})\mathbf{A}_{n'\mathbf{k}'}^{*}(\mathbf{r})=V\delta_{nn'}\delta(\mathbf{k}-\mathbf{k}'),\label{math:normalization}\\
 &  & \int d^{3}\mathbf{kA}_{n\mathbf{k}}(\mathbf{r})\mathbf{A}_{n\mathbf{k}}^{*}(\mathbf{r}')=\mathcal{I}_{\varepsilon_{\bot}}\delta(\mathbf{r}-\mathbf{r}'),\label{math:closure}\end{eqnarray}
where $\omega_{n\mathbf{k}}$ is the Bloch eigenfrequency, $V$ is
the volume of the unit cell of the crystal, $^{*}$ denotes the complex
conjugate and $\mathcal{I}_{\varepsilon_{\bot}}$ is the identity
operator on the subset of the $\varepsilon$-transverse vector functions
as defined in \cite{glauber91}. The Bloch eigenvector $\mathbf{A}_{n\mathbf{k}}(\mathbf{r})$
obeys the gauge condition $\nabla\cdot[\varepsilon(\mathbf{r})\mathbf{A}_{n\mathbf{k}}(\mathbf{r})]=0$
and are therefore transverse with respect to this gauge. Equations
(\ref{math:normalization}-\ref{math:closure}) ensure that the eigenvectors
$\mathbf{A}_{n\mathbf{k}}(\mathbf{r})$ form a complete set of orthonormal
$\varepsilon$-transverse functions. Here any vector that satisfies
the $\varepsilon$-transverse gauge condition (\ref{math:gauge})
is called {}``$\varepsilon$-transverse'' \cite{dowling92}.

\begin{widetext}The amplitude coefficients $C_{n\mathbf{k}}(t)$
can be easily obtained from the wave equation (\ref{math:waveEQ-A}).
Substituting (\ref{math:generalField}) into the wave equation (\ref{math:waveEQ-A})
and using the homogeneous wave equation (\ref{math:waveEQ-Ank}),
one obtains\[
\sum_{n}\int_{BZ}d^{3}\mathbf{k}_{n}\left(\frac{\partial^{2}C_{n\mathbf{k}}(t)}{\partial t^{2}}+\omega_{n\mathbf{k}}^{2}C_{n\mathbf{k}}(t)\right)\varepsilon(\mathbf{r})\mathbf{A}_{n\mathbf{k}}(\mathbf{r})=4\pi c\mathbf{J}(\mathbf{r},t)\]
\end{widetext}Then taking the inner product between every term of
this equation and an eigenvector $\mathbf{A}_{n'\mathbf{k}'}(\mathbf{r})$,
i.e., multiplying by $\mathbf{A}_{n'\mathbf{k}'}^{*}(\mathbf{r})$
and integrating over the unit cell of the crystal, one finally obtains
the differential equation for the amplitude coefficients $C_{n\mathbf{k}}(t)$\[
\frac{\partial^{2}C_{n\mathbf{k}}(t)}{\partial t^{2}}+\omega_{n\mathbf{k}}^{2}C_{n\mathbf{k}}(t)=-i\frac{4\pi c\omega_{0}}{V}\left(\mathbf{A}_{n\mathbf{k}}^{*}(\mathbf{r}_{0})\cdot\mathbf{d}\right)e^{-i\omega_{0}t},\]
where the orthogonality of the eigenvectors (\ref{math:normalization})
and a specific form of the source term (\ref{math:sourceTerm}) were
taken into account. Then assuming initial conditions $C_{n\mathbf{k}}(0)=0$,
one has the following solution of this differential equation \begin{equation}
C_{n\mathbf{k}}(t)=-i\frac{4\pi c\omega_{0}}{V}\frac{\left(\mathbf{A}_{n\mathbf{k}}^{*}(\mathbf{r}_{0})\cdot\mathbf{d}\right)}{\left(\omega_{n\mathbf{k}}^{2}-\omega_{0}^{2}\right)}e^{-i\omega_{0}t}.\label{math:Cnk}\end{equation}
Finally, the electromagnetic field radiated by a point dipole located
at $\mathbf{r}_{0}$ can be represented in terms of Bloch normal modes
as: \begin{eqnarray}
\mathbf{A}(\mathbf{r},t) & = & -i\frac{4\pi c\omega_{0}}{V}\sum_{n}\int_{BZ}d^{3}\mathbf{k}_{n}\frac{\left(\mathbf{a}_{n\mathbf{k}}^{*}(\mathbf{r}_{0})\cdot\mathbf{d}\right)}{\left(\omega_{n\mathbf{k}}^{2}-\omega_{0}^{2}\right)}\nonumber \\
 & \times & \mathbf{a}_{n\mathbf{k}}(\mathbf{r})e^{i\mathbf{k}_{n}\cdot(\mathbf{r}-\mathbf{r}_{0})}e^{-i\omega_{0}t},\label{math:fieldDipole}\end{eqnarray}
where the Bloch theorem, $\mathbf{A}_{n\mathbf{k}}(\mathbf{r})=\mathbf{a}_{n\mathbf{k}}(\mathbf{r})e^{i\mathbf{k}_{n}\cdot\mathbf{r}}$,
have been used.

The integrand in (\ref{math:fieldDipole}) has a pole at $\omega_{n\mathbf{k}}^{2}=\omega_{0}^{2}$,
and the integral is singular. This is a typical behavior for any resonance
system, where dissipation is neglected. The standard way to regularize
the integral is to add a small imaginary part to $\omega_{0}^{2}$.
The result of the integration then becomes dependent on the sign of
this imaginary part. The criterion for determining the sign will be
discussed below. A regularized integral (\ref{math:fieldDipole})
reads\begin{eqnarray}
\mathbf{A}(\mathbf{r},t) & = & -i\frac{4\pi c\omega_{0}}{V}\sum_{n}\int_{BZ}d^{3}\mathbf{k}_{n}\frac{\left(\mathbf{a}_{n\mathbf{k}}^{*}(\mathbf{r}_{0})\cdot\mathbf{d}\right)}{\left(\omega_{n\mathbf{k}}^{2}-\omega_{0}^{2}-i\gamma\right)}\nonumber \\
 & \times & \mathbf{a}_{n\mathbf{k}}(\mathbf{r})e^{i\mathbf{k}_{n}\cdot(\mathbf{r}-\mathbf{r}_{0})}e^{-i\omega_{0}t}.\label{math:fieldRegularized}\end{eqnarray}

\section{\label{sec:emission_rate}Spontaneous emission rate}

A light source situated in an inhomogeneous medium is immersed in
its own electric field emitted at an earlier time and reflected from
inhomogeneities in the medium. By conservation of energy, the decay
rate at which energy is radiated is equal to the rate at which the
charge distribution of the source does work on the surrounding electromagnetic
field. For an arbitrary current density $\mathbf{J}$, the radiated
power is given by \cite{bookJackson}:\begin{equation}
P(t)=-\int_{V}d^{3}\mathbf{rJ}(\mathbf{r},t)\cdot\mathbf{E}(\mathbf{r},t),\label{math:rate}\end{equation}
where $V$ is a volume containing a current density source $\mathbf{J}$
and it is related to spontaneous emission rate via $\Gamma=P/\hbar\omega_{0}$
\cite{ford84}. For the time-averaged radiated power, one has \begin{equation}
P=-\frac{1}{2}\mathrm{Re}\left[\int_{V}d^{3}\mathbf{rJ}^{*}(\mathbf{r},t)\cdot\mathbf{E}(\mathbf{r},t)\right],\end{equation}
or, specializing to a point dipole (\ref{math:sourceTerm}) \begin{equation}
P=\frac{\omega_{0}}{2}\mathrm{Im}\left[\mathbf{d}^{*}\cdot\mathbf{E}(\mathbf{r}_{0})\right].\label{math:powerP}\end{equation}
A well known interpretation of the emission rate modification follows
from this relation. Emission rate is modified due to the dipole interaction
with the out-of-phase part of the radiation reaction field \cite{sipe81,wylie84}.

\begin{figure}[tb]
\begin{center}\includegraphics[%
  width=0.80\columnwidth]{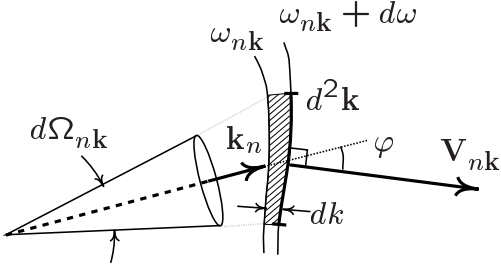}\end{center}

\caption{Diagram showing the relations between \textbf{k}-space and coordinate
space quantities. Iso-frequency contours for frequencies $\omega_{n\mathbf{k}}$
and $\omega_{n\mathbf{k}}+d\omega$ are presented. \label{fig:r-k-space}}
\end{figure}

Consider an excited molecule or atom at a position $\mathbf{r}_{0}$
in a photonic crystal. Assuming that the presence of the molecule
does not change the band structure of the crystal, the only possible
mode it can emit in, is an eigenmode of the photonic crystal. In the
classical approach, the molecule is modeled by a point dipole $\mathbf{d}$
(\ref{math:sourceTerm}). The radiation reaction field will be given
by the normal mode expansion (\ref{math:fieldRegularized}), which
is valid for any point $\mathbf{r}$ in the crystal, which is distinct
from (but as close as required to) the dipole location $\mathbf{r}_{0}$.
Such a choice of the radiation reaction field corresponds to the Weisskopf-Wigner
approximation \cite{glauber91,busch00}. Then the radiated power (emission
rate) (\ref{math:powerP}) of a classical dipole in a photonic crystal
is given by: \begin{equation}
P=\mathrm{Im}\left[\frac{2\pi\omega_{0}^{3}}{V}\sum_{n}\int_{BZ}d^{3}\mathbf{k}_{n}\frac{\left|\mathbf{A}_{n\mathbf{k}}(\mathbf{r}_{0})\cdot\mathbf{d}\right|^{2}}{\left(\omega_{n\mathbf{k}}^{2}-\omega_{0}^{2}-i\gamma\right)}\right].\label{math:powerP1}\end{equation}
where equation (\ref{math:EviaA}), relating $\mathbf{E}=-(1/c)\partial\mathbf{A}/\partial t$,
was used. This integral can be converted to a two-dimensional integral
over the iso-frequency surface in \textbf{k}-space. With the aid of
the integral representation\begin{equation}
\frac{1}{x-i\gamma}=-\frac{1}{i}\int_{0}^{\infty}d\tau e^{-ix\tau-\gamma\tau},\label{math:x-iy}\end{equation}
one can transform the integral (\ref{math:powerP1}) to the form:
\begin{eqnarray}
P & = & \mathrm{Im}\left[i\frac{2\pi\omega_{0}^{3}}{V}\sum_{n}\int_{BZ}d^{3}\mathbf{k}_{n}\left|\mathbf{A}_{n\mathbf{k}}(\mathbf{r}_{0})\cdot\mathbf{d}\right|^{2}\right.\nonumber \\
 & \times & \left.\int_{0}^{\infty}dte^{i(\omega_{0}^{2}-\omega_{n\mathbf{k}}^{2})t}\right].\label{math:powerP2}\end{eqnarray}
Now, making use of the time-reversal symmetry of the Maxwell's equations,
which for a periodic medium implies that $\omega_{n,\mathbf{k}}=\omega_{n,-\mathbf{k}}$
\cite{bookJDJ}, one can rewrite the time integral in (\ref{math:powerP2})
to obtain\[
\int_{0}^{\infty}dte^{i(\omega_{0}^{2}-\omega_{n\mathbf{k}}^{2})t}=\frac{1}{2}\int_{-\infty}^{\infty}dte^{i(\omega_{0}^{2}-\omega_{n\mathbf{k}}^{2})t}=\pi\delta(\omega_{0}^{2}-\omega_{n\mathbf{k}}^{2}).\]
 Then, changing the integration variable to the eigenfrequency $\omega_{n\mathbf{k}}$
by use of the relations $\left|\nabla_{\mathbf{k}}\omega_{n\mathbf{k}}\right|dk=d\omega_{n\mathbf{k}}$
and $d^{3}\mathbf{k}_{n}=dkd^{2}\mathbf{k}_{n}$, where $d^{2}\mathbf{k}_{n}$
is an element of the iso-frequency surface $\omega_{n\mathbf{k}}=\omega_{0}$
(Fig.~\ref{fig:r-k-space}), equation (\ref{math:powerP2}) can be
transformed to:\[
P=\frac{2\pi^{2}\omega_{0}^{3}}{V}\sum_{n}\int d^{2}\mathbf{k}_{n}\int d\omega_{n\mathbf{k}}\frac{\left|\mathbf{A}_{n\mathbf{k}}(\mathbf{r}_{0})\cdot\mathbf{d}\right|^{2}}{\left|\nabla_{\mathbf{k}}\omega_{n\mathbf{k}}\right|}\delta(\omega_{0}^{2}-\omega_{n\mathbf{k}}^{2}),\]
where one can carry out the integration over the eigenfrequency $\omega_{n\mathbf{k}}$
to obtain finally\begin{equation}
P=\frac{\pi^{2}\omega_{0}^{2}}{V}\sum_{n}\int d^{2}\mathbf{k}_{n}\frac{\left|\mathbf{A}_{n\mathbf{k}}(\mathbf{r}_{0})\cdot\mathbf{d}\right|^{2}}{\left|\mathbf{V}_{n\mathbf{k}}\right|},\label{math:powerFinal-Kspace}\end{equation}
where, $\mathbf{V}_{n\mathbf{k}}=\nabla_{\mathbf{k}}\omega_{n\mathbf{k}}$,
the group velocity of the eigenwave $(n,\mathbf{k})$ is introduced.

Formula (\ref{math:powerFinal-Kspace}) gives the total time-averaged
radiated power of a dipole situated inside a photonic crystal \cite{dowling92,sakoda96}.
This result agrees with fully quantum electrodynamical result for
spontaneous emission rate of a two-level atom within the Weisskopf-Wigner
approximation \cite{glauber91}.

\section{\label{sec:asymptotic_far_field}Asymptotic form of dipole field}

In this section, a radiating dipole field is analyzed in the radiation
zone. For that, an asymptotic form of the integral (\ref{math:fieldRegularized})
is evaluated and analyzed. In what follows, an asymptotic analysis
of the Green's function developed by Maradudin \cite{maradudin64}
for the phonon scattering problem is used.

Using the integral representation (\ref{math:x-iy}) one can rewrite
(\ref{math:fieldRegularized}) as\begin{eqnarray}
\mathbf{A}(\mathbf{r}) & = & \frac{4\pi c\omega_{0}}{V}\sum_{n}\int_{BZ}d^{3}\mathbf{k}_{n}\nonumber \\
 & \times & \int_{0}^{\infty}d\tau\left(\mathbf{a}_{n\mathbf{k}}^{*}(\mathbf{r}_{0})\cdot\mathbf{d}\right)\mathbf{a}_{n\mathbf{k}}(\mathbf{r})e^{iF_{n\mathbf{k}}(\tau)},\label{math:fieldFnk}\end{eqnarray}
where\begin{equation}
F_{n\mathbf{k}}(\tau)=\mathbf{k}_{n}\cdot(\mathbf{r}-\mathbf{r}_{0})-\tau(\omega_{n\mathbf{k}}^{2}-\omega_{0}^{2})\label{math:Fnk}\end{equation}
and a limit $\gamma\rightarrow0$ was taken.

In a typical experiment $\mathbf{\left|x\right|}=\left|\mathbf{r}-\mathbf{r}_{0}\right|\gg\lambda$,
where $\lambda$ is the wavelength of the electromagnetic wave. For
large $\left|\mathbf{x}\right|$ an exponential function in the integral
(\ref{math:fieldFnk}) will oscillate very rapidly and one can use
the method of stationary phase to evaluate the integral.

The principal contribution to the integral comes from the neighborhood
of those points in $\tau$- and $\mathbf{k}$-space where the variation
of \textbf{$F_{n\mathbf{k}}(\tau)$} is the smallest. This means that
one can set the gradient of the function \textbf{$F_{n\mathbf{k}}(\tau)$}
in \textbf{k}-space equal to zero as well as the derivative of the
function with respect to $\tau$. This gives the conditions \begin{eqnarray}
 &  & \frac{\partial F_{n\mathbf{k}}}{\partial\tau}=\omega_{n\mathbf{k}}^{2}-\omega_{0}^{2}=0,\label{math:stationaryOmega}\\
 &  & \nabla_{\mathbf{k}}F_{n\mathbf{k}}=\mathbf{x}-\tau\nabla_{\mathbf{k}}\omega_{n\mathbf{k}}^{2}=0.\label{math:stationaryK}\end{eqnarray}
Equations (\ref{math:stationaryOmega}-\ref{math:stationaryK}) determine
the values of $\tau$ and $\mathbf{k}_{n}$ around which the principal
contributions to the integral (\ref{math:fieldFnk}) arise. These
points are called stationary points. Further, the stationary points
are denoted by $\tau_{\nu}$ and $\mathbf{k}_{n}^{\nu}$. Assuming
that value of the eigenvector $\mathbf{a}_{n\mathbf{k}}(\mathbf{r})$
is approximately constant $\mathbf{a}_{n\mathbf{k}}(\mathbf{r})\approx\mathbf{a}_{n\mathbf{k}}^{\nu}(\mathbf{r})$
for $\tau$ close to $\tau_{\nu}$ and for the wave vectors close
to $\mathbf{k}_{n}^{\nu}$, the integral (\ref{math:fieldFnk}) is
reduced to the sum of the integrals in the vicinities of the stationary
points $(\tau_{\nu},\mathbf{k}_{n}^{\nu}$) \cite{maradudin64,maris1983}\begin{eqnarray}
\mathbf{A}(\mathbf{r}) & \approx & \frac{4\pi c\omega_{0}}{V}\sum_{n}\sum_{\nu}\left(\mathbf{a}_{n\mathbf{k}}^{\nu*}(\mathbf{r}_{0})\cdot\mathbf{d}\right)\mathbf{a}_{n\mathbf{k}}^{\nu}(\mathbf{r})\nonumber \\
 & \times & \int_{\mathbf{k}_{n}^{\nu}}d^{3}\mathbf{k}_{n}\int_{\tau_{\nu}}d\tau e^{iF_{n\mathbf{k}}(\tau)},\label{math:FnkAsymtotic}\end{eqnarray}
Here an extra summation is over all possible solutions of Eqs. (\ref{math:stationaryOmega}-\ref{math:stationaryK}).

Due to Eq. (\ref{math:stationaryOmega}) the principal contribution
to the asymptotic behavior of $\mathbf{A}(\mathbf{r})$ comes from
the iso-frequency surface in $\mathbf{k}$-space defined by \textbf{$\omega_{n\mathbf{k}}^{2}=\omega_{0}^{2}$}
or equivalently defined by $\omega_{n\mathbf{k}}=\omega_{0}$ (eigenfrequency
$\omega_{n\mathbf{k}}$ is positive and real). At the same time, due
to Eq. (\ref{math:stationaryK}) the portion of the iso-frequency
surface $\omega_{n\mathbf{k}}=\omega_{0}$, which contributes to the
asymptotic field, is the portion near the point on this surface where
the gradient $\nabla_{\mathbf{k}}\omega_{n\mathbf{k}}^{2}$ is parallel
to $\mathbf{x}$. One can express the latter condition in an alternative
fashion. Equation (\ref{math:stationaryK}) can be simplified as:
\[
\mathbf{x}=2\tau\omega_{n\mathbf{k}}\mathbf{V}_{n\mathbf{k}},\]
where $\mathbf{V}_{n\mathbf{k}}=\nabla_{\mathbf{k}}\omega_{n\mathbf{k}}$
is the group velocity of the eigenwave $(n,\mathbf{k})$. So, equation
(\ref{math:stationaryK}) just says that the principal contribution
to the asymptotic behavior of the field $\mathbf{A}(\mathbf{r})$
at large $\left|\mathbf{x}\right|=\left|\mathbf{r}-\mathbf{r}_{0}\right|\gg\lambda$
comes from the neighborhood of the points $\mathbf{k}_{n}^{\nu}$
on the iso-frequency surface $\omega_{n\mathbf{k}}=\omega_{0}$ at
which the eigenwave group velocity is collinear to observation direction
$\mathbf{x}$. Since $\tau$ is positive by definition (\ref{math:x-iy}),
$\mathbf{V}_{n\mathbf{k}}^{\nu}$ and $\mathbf{x}$ should not only
be collinear, but should point in the same direction as well, i. e.,
$\mathbf{x}\cdot\mathbf{V}_{n\mathbf{k}}^{\nu}>0$.

Assuming that the major contribution comes from the regions near the
stationary points, one makes a little error by extending the integration
in (\ref{math:FnkAsymtotic}) over all space\begin{eqnarray}
\mathbf{A}(\mathbf{r}) & \approx & \frac{4\pi c\omega_{0}}{V}\sum_{n}\sum_{\nu}\left(\mathbf{a}_{n\mathbf{k}}^{\nu*}(\mathbf{r}_{0})\cdot\mathbf{d}\right)\mathbf{a}_{n\mathbf{k}}^{\nu}(\mathbf{r})\nonumber \\
 & \times & \int_{-\infty}^{\infty}d^{3}\mathbf{k}_{n}\int_{-\infty}^{\infty}d\tau e^{iF_{n\mathbf{k}}(\tau)}.\label{math:FnkAsymtoticINF}\end{eqnarray}
Then, the integral over $\tau$ is simply given by Dirac $\delta$-function:\[
\int_{-\infty}^{\infty}d\tau e^{i\tau(\omega_{0}^{2}-\omega_{n\mathbf{k}}^{2})}=2\pi\delta(\omega_{0}^{2}-\omega_{n\mathbf{k}}^{2})\]
and one can further convert the volume integration in $\mathbf{k}$-space
to an integral over the iso-frequency surface $\omega_{n\mathbf{k}}=\omega_{0}$.
In fact, by using the relations $\left|\nabla_{\mathbf{k}}\omega_{n\mathbf{k}}\right|dk=d\omega_{n\mathbf{k}}$
and $d^{3}\mathbf{k}=dkd^{2}\mathbf{k}$, and integrating over the
eigenfrequency $\omega_{n\mathbf{k}}$, the volume integration over
$\mathbf{k}$ transforms to:\[
\int_{-\infty}^{\infty}d^{3}\mathbf{k}_{n}e^{i\mathbf{k}_{n}(\mathbf{r}-\mathbf{r}_{0})}\delta(\omega_{0}^{2}-\omega_{n\mathbf{k}}^{2})=\oint_{-\infty}^{\infty}d^{2}\mathbf{k}_{n}\frac{\pi}{\omega_{0}}\frac{e^{i\mathbf{k}_{n}\cdot(\mathbf{r}-\mathbf{r}_{0})}}{\left|\mathbf{V}_{n\mathbf{k}}\right|},\]
where $\mathbf{V}_{n\mathbf{k}}=\nabla_{\mathbf{k}}\omega_{n\mathbf{k}}$
is the group velocity of the eigenwave $(n,\mathbf{k})$. So, the
asymptotic form of the field $\mathbf{A}(\mathbf{r})$ is given finally
by:\begin{eqnarray}
\mathbf{A}(\mathbf{r}) & \approx & \frac{4\pi^{2}c}{V}\sum_{n}\sum_{\nu}\frac{\left(\mathbf{a}_{n\mathbf{k}}^{\nu*}(\mathbf{r}_{0})\cdot\mathbf{d}\right)\mathbf{a}_{n\mathbf{k}}^{\nu}(\mathbf{r})}{\left|\mathbf{V}_{n\mathbf{k}}^{\nu}\right|}\nonumber \\
 & \times & \oint_{-\infty}^{\infty}d^{2}\mathbf{k}_{n}e^{i\mathbf{k}_{n}\cdot(\mathbf{r}-\mathbf{r}_{0})},\label{math:FnkAsymptotic2D}\end{eqnarray}
 where the comparatively slowly varying function $\mathbf{V}_{n\mathbf{k}}$
was replaced by its value at stationary point $\mathbf{k}_{n}^{\nu}$
and was taken outside the integral over $\mathbf{k}$.

To evaluate the integrals in Eq. (\ref{math:FnkAsymptotic2D}) the
analysis of the form of the iso-frequency surface in the vicinity
of one of the stationary points, $\mathbf{k}_{n}^{\nu}$, should be
done. It is convenient to introduce the local curvilinear coordinates
$\xi_{i}$ with the origin at the stationary point and with one of
the coordinate aligned perpendicular to the iso-frequency surface,
e.g., $\xi_{3}$. One can expand function $h\left(\xi_{1},\xi_{2}\right)=\mathbf{k}_{n}\cdot\widehat{\mathbf{x}}$
near the stationary point as:

\begin{eqnarray}
h\left(\xi_{1},\xi_{2}\right) & = & \mathbf{k}_{n}^{\nu}\cdot\widehat{\mathbf{x}}+\frac{1}{2}\sum_{i,j=1}^{2}\alpha_{ij}^{\nu}\xi_{i}\xi_{j}\nonumber \\
 & + & \frac{1}{6}\sum_{i,j,k=1}^{2}\beta_{ijk}^{\nu}\xi_{i}\xi_{j}\xi_{k}+O\left(\xi_{1},\xi_{2}\right)^{4},\label{math:surfaceExpansion}\end{eqnarray}
where\[
\alpha_{ij}^{\nu}=\left(\frac{\partial^{2}h}{\partial\xi_{i}\partial\xi_{j}}\right)_{\nu},\qquad\beta_{ijk}^{\nu}=\left(\frac{\partial^{3}h}{\partial\xi_{i}\partial\xi_{j}\partial\xi_{k}}\right)_{\nu}\]
and $\widehat{\mathbf{x}}$ is a unit vector in the observation direction.
All derivatives are evaluated at the stationary point $\mathbf{k}_{n}^{\nu}$.

\begin{figure}[tb]
\begin{center}\includegraphics[%
  width=0.95\columnwidth]{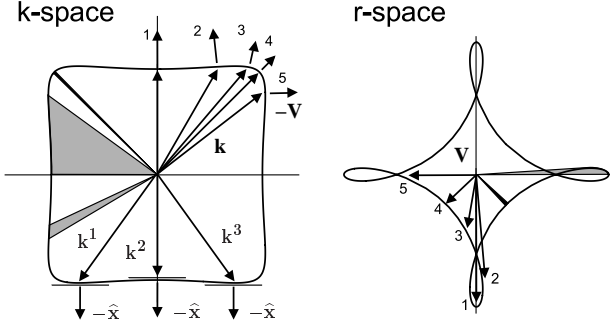}\end{center}

\caption{\label{fig:folding}Iso-frequency and wave contours. Left: The central
region of the iso-frequency contour for normalized frequency $\Omega=\omega d/2\pi c=d/\lambda=0.569$
of an infinite square lattice 2D photonic crystal made out of dielectric
rods placed in vacuum. Rods have the optical index 2.9 and radius
$r=0.15d$, where $d$ is the period of the lattice (see Sec. \ref{sec:example_2D_PhC}
for details). The stationary points, $\mathbf{k}^{1}$, $\mathbf{k}^{2}$
and $\mathbf{k}^{3}$, corresponding to the same observation direction
$\widehat{\mathbf{x}}$ are indicated. Right: Corresponding wave contour
with folds. The shaded and black regions show how two equal solid
angle sections in coordinate space (right) map to widely varying solid
angle sections in \textbf{k}-space (left). The wave and group velocity
vectors with numbers illustrate the folds formation of the wave contour.}
\end{figure}

The result of the integration in (\ref{math:FnkAsymptotic2D}) depends
on the local topology of the iso-frequency surface near the stationary
point. One can generally classify the local topology of the surface
by its Gaussian curvature. The Gaussian curvature $K$ is the product
of the two principal curvatures (inverse radii, $K_{1}$ and $K_{2}$)
at a point on the surface, i.e., $K=K_{1}K_{2}$. The points on an
iso-frequency surface can be elliptical, hyperbolic and parabolic.
If the Gaussian curvature $K>0$, the corresponding point on the iso-frequency
surface is called elliptical, and if $K<0$ it is called hyperbolic.
For a complex surface, such as the iso-frequency surface in figure
\ref{fig:folding}-left, the regions with positive and negative Gaussian
curvature alternate. The surface is parabolic at the borders between
regions with curvatures of opposite signs, e.g, convex and saddle.
The lines along which the curvature changes its sign are called parabolic
lines. The Gaussian curvature at a parabolic point is equal to zero.

Further, the analysis of the asymptotic form of the integral (\ref{math:FnkAsymptotic2D})
is undertaken, when the stationary points are elliptical or hyperbolic.
Then in the close vicinity of such a stationary point the following
expansion holds:\begin{equation}
h\left(\xi_{1},\xi_{2}\right)=\mathbf{k}_{n}^{\nu}\cdot\widehat{\mathbf{x}}+\frac{1}{2}\sum_{i,j=1}^{2}\alpha_{ij}^{\nu}\xi_{i}\xi_{j},\end{equation}
where only quadratic terms in the expansion (\ref{math:surfaceExpansion})
were kept. By choosing the orientation of the local coordinates $\xi_{1}$
and $\xi_{2}$ along the main directions of the surface curvature
at that point $\mathbf{k}_{n}=\mathbf{k}_{n}^{\nu}$, one can diagonalize
the matrix $\alpha_{ij}^{\nu}$. Then\begin{equation}
h\left(\xi_{1},\xi_{2}\right)=\mathbf{k}_{n}^{\nu}\cdot\widehat{\mathbf{x}}+\frac{1}{2}\left(\alpha_{1}^{\nu}\xi_{1}^{2}+\alpha_{2}^{\nu}\xi_{2}^{2}\right),\,\alpha_{1}^{\nu}=\alpha_{11}^{\nu},\:\alpha_{2}^{\nu}=\alpha_{22}^{\nu}.\label{math:hElliptical}\end{equation}
With such a choice of local coordinates in \textbf{k}-space, the product
$K_{n\mathbf{k}}^{\nu}=\alpha_{1}^{\nu}\alpha_{2}^{\nu}$ determines
the Gaussian curvature of the iso-frequency surface at the stationary
point $\mathbf{k}_{n}=\mathbf{k}_{n}^{\nu}$. 

\begin{widetext}Using expansion (\ref{math:hElliptical}) the asymptotic
form of the field (\ref{math:FnkAsymptotic2D}) is now given by:\begin{equation}
\mathbf{A}(\mathbf{r})\approx\frac{4\pi^{2}c}{V}\sum_{n}\sum_{\nu}\frac{\left(\mathbf{a}_{n\mathbf{k}}^{\nu*}(\mathbf{r}_{0})\cdot\mathbf{d}\right)\mathbf{a}_{n\mathbf{k}}^{\nu}(\mathbf{r})}{\left|\mathbf{V}_{n\mathbf{k}}^{\nu}\right|}e^{i\mathbf{k}_{n}^{\nu}\cdot\mathbf{x}}\oint_{-\infty}^{\infty}d\xi_{1}d\xi_{2}\exp\left(\frac{i\left|\mathbf{x}\right|}{2}\left(\alpha_{1}^{\nu}\xi_{1}^{2}+\alpha_{2}^{\nu}\xi_{2}^{2}\right)\right).\label{math:FnElliptical}\end{equation}
The integral in Eq. (\ref{math:FnElliptical}) is calculated simply
to be \begin{equation}
\int_{-\infty}^{\infty}d\xi\exp\left(i\frac{x\alpha}{2}\xi^{2}\right)=\sqrt{\frac{2\pi}{x\left|\alpha\right|}}\exp\left(-\frac{i\pi}{4}\mathrm{sign}\left(\alpha\right)\right)\label{math:integral1}\end{equation}
and an asymptotic form of the vector potential (\ref{math:fieldDipole})
at the position $\mathbf{r}$ far from the dipole is given by \begin{equation}
\mathbf{A}(\mathbf{r},t)\approx\sum_{n}\sum_{\nu}\exp\left(-i\left(\omega_{0}t+\frac{\pi}{4}\left(\mathrm{sign}(\alpha_{1}^{\nu})+\mathrm{sign}(\alpha_{2}^{\nu})\right)\right)\right)\frac{c}{V}\frac{\left(\mathbf{A}_{n\mathbf{k}}^{\nu*}(\mathbf{r}_{0})\cdot\mathbf{d}\right)\mathbf{A}_{n\mathbf{k}}^{\nu}(\mathbf{r})}{\left|\mathbf{V}_{n\mathbf{k}}^{\nu}\right|}\frac{8\pi^{3}}{\left|K_{n\mathbf{k}}^{\nu}\right|^{1/2}\left|\mathbf{r}-\mathbf{r}_{0}\right|}\label{math:FnAsymGeometrical}\end{equation}
where $\mathbf{A}_{n\mathbf{k}}^{\nu}(\mathbf{r})=\mathbf{a}_{n\mathbf{k}}^{\nu}(\mathbf{r})e^{i\mathbf{k}_{n}^{\nu}\cdot\mathbf{r}}$
and summation is over all stationary points with $\mathbf{x}\cdot\mathbf{V}_{n\mathbf{k}}^{\nu}>0$.\end{widetext}

According to the Eq. (\ref{math:FnAsymGeometrical}) the electromagnetic
field inside photonic crystal represents a superposition of several
diverging waves, the number of which equals the number of stationary
phase points on the iso-frequency surface $\omega_{n\mathbf{k}}=\omega_{0}$
(Fig.~\ref{fig:folding}-left). Each of these waves has its own shape
and its own propagation velocity. One comment is important here, the
asymptotic expansion (\ref{math:FnAsymGeometrical}) describes an
outgoing wave ($\mathbf{k}_{n}^{\nu}\cdot\mathbf{x}>0$) only if the
corresponding group velocity is an outward normal to the iso-frequency
surface $\omega_{n\mathbf{k}}=\omega_{0}$ at point $\mathbf{k}_{n}^{\nu}$.
It can happen, however, that the group velocity becomes an inward
normal for some frequencies and regions of \textbf{k}-space (Fig.~\ref{fig:folding}-left).
In such a case the dot product $\mathbf{k}_{n}^{\nu}\cdot\mathbf{x}$
is not positive in the asymptotic expansion (\ref{math:FnAsymGeometrical})
and the expansion describes incoming waves. In such a situation, one
should change the sign of the small imaginary part $\gamma$ in regularized
equation (\ref{math:fieldRegularized}) \cite{maradudin64}: \begin{eqnarray}
\mathbf{A}(\mathbf{r}) & = & -i\frac{4\pi c\omega_{0}}{V}\sum_{n}\int_{BZ}d^{3}\mathbf{k}_{n}\frac{\left(\mathbf{a}_{n\mathbf{k}}^{*}(\mathbf{r}_{0})\cdot\mathbf{d}\right)}{\left(\omega_{n\mathbf{k}}^{2}-\omega_{0}^{2}+i\gamma\right)}\nonumber \\
 & \times & \mathbf{a}_{n\mathbf{k}}(\mathbf{r})e^{i\mathbf{k}_{n}\cdot(\mathbf{r}-\mathbf{r}_{0})}\label{math:fieldRegularized+}\end{eqnarray}
and proceed as it has been describe above (\ref{math:fieldFnk}-\ref{math:FnAsymGeometrical}),
but using the integral representation\begin{equation}
\frac{1}{x+i\gamma}=\frac{1}{i}\int_{0}^{\infty}d\tau e^{ix\tau-\gamma\tau}\label{math:x+iy}\end{equation}
instead of (\ref{math:x-iy}).

\section{\label{sec:angular_distribution_of_power}Angular distribution of
radiated power}

In this section, the angular dependence of the dipole radiated power
(\ref{math:powerFinal-Kspace}) is introduced.

Using the definition of the solid angle, $d\Omega_{n\mathbf{k}}=d^{2}\mathbf{k}\cos\varphi/\left|\mathbf{k}_{n}\right|^{2}$,
where $d\Omega_{n\mathbf{k}}$ is the solid angle subtended by the
surface element $d^{2}\mathbf{k}_{n}$, $\varphi$ is the angle between
the wave vector $\mathbf{k}_{n}$ and the group velocity $\mathbf{V}_{n\mathbf{k}}=\nabla_{\mathbf{k}}\omega_{n\mathbf{k}}$
(Fig.~\ref{fig:r-k-space}), on changing the integration variables,
one can modify equation (\ref{math:powerFinal-Kspace}) to the form\begin{equation}
P=\sum_{n}\int_{0}^{4\pi}d\Omega_{n\mathbf{k}}\left(\frac{\pi^{2}\omega_{0}^{2}}{V}\frac{\left|\mathbf{A}_{n\mathbf{k}}(\mathbf{r}_{0})\cdot\mathbf{d}\right|^{2}}{\left|\mathbf{V}_{n\mathbf{k}}\right|}\frac{\left|\mathbf{k}_{n}\right|^{2}}{\cos\varphi}\right),\label{math:powerOmegaNK}\end{equation}
where the function enclosed in the brackets defines the radiated power
of the dipole per solid angle in \textbf{k}-space\begin{equation}
\frac{dP}{d\Omega_{n\mathbf{k}}}=\frac{\pi^{2}\omega_{0}^{2}}{V}\frac{\left|\mathbf{A}_{n\mathbf{k}}(\mathbf{r}_{0})\cdot\mathbf{d}\right|^{2}}{\left|\mathbf{V}_{n\mathbf{k}}\right|}\frac{\left|\mathbf{k}_{n}\right|^{2}}{\cos\varphi}.\label{math:angularKspace}\end{equation}
To derive the angular distribution of radiated power in the coordinate
space, one should change the integration variables in (\ref{math:powerOmegaNK})
from the \textbf{k}-space to the coordinate space. 

The \textbf{k}-space distribution of the radiated power (\ref{math:angularKspace})
is a function of the \textbf{k}-space direction, given by the polar,
$\theta_{n\mathbf{k}}$, and azimuthal, $\phi_{n\mathbf{k}}$, angles
of the wave vector $\mathbf{k}_{n}$. The direction of energy propagation
in a non-absorbing periodic medium coincides with the group velocity
direction \cite{bookYeh88}. Whereas the coordinate space angular
dependence of the radiated power is given by the corresponding group
velocity direction in the coordinate space $(\theta,\phi)$. Here
$\theta$ and $\phi$ are the polar and azimuthal angles of the group
velocity in coordinate space. The \textbf{k}-space to the coordinate
space transformation may be expressed formally as \begin{eqnarray}
 &  & \cos\theta=f(\cos\theta_{n\mathbf{k}},\phi_{n\mathbf{k}}),\label{math:mappingTheta}\\
 &  & \phi=g(\cos\theta_{n\mathbf{k}},\phi_{n\mathbf{k}}),\label{math:mappingPhi}\end{eqnarray}
where the functions $f$ and $g$ are determined from the components
of the group velocity vector $\mathbf{V}_{n\mathbf{k}}^{\nu}\parallel\widehat{\mathbf{x}}$,
where $\widehat{\mathbf{x}}$ is a unit vector in the observation
direction. The Jacobian of the transformation (\ref{math:mappingTheta}-\ref{math:mappingPhi})
\begin{equation}
J_{n\mathbf{k}}=\frac{\partial f}{\partial\cos\theta_{n\mathbf{k}}}\frac{\partial g}{\partial\phi_{n\mathbf{k}}}-\frac{\partial f}{\partial\phi_{n\mathbf{k}}}\frac{\partial g}{\partial\cos\theta_{n\mathbf{k}}}\label{math:jacobian}\end{equation}
relates a small solid angle in the coordinate space with the corresponding
solid angle in \textbf{k}-space via \begin{equation}
d\Omega=d(\cos\theta)d\phi=J_{n\mathbf{k}}d(\cos\theta_{n\mathbf{k}})d\phi_{n\mathbf{k}}=J_{n\mathbf{k}}d\Omega_{n\mathbf{k}}.\label{math:dOmega:dOmegak}\end{equation}
 According to the results presented in the Section \ref{sec:asymptotic_far_field},
different wave vectors can result in the group velocity with same
direction in coordinate space. That means that the following equation\[
d\Omega_{n\mathbf{k}}^{\nu}=\frac{1}{J_{n\mathbf{k}}^{\nu}}d\Omega\]
should hold for each stationary wave vector, which satisfies $\widehat{\mathbf{x}}\cdot\mathbf{V}_{n\mathbf{k}}^{\nu}>0$.
Changing the integration variables in (\ref{math:powerOmegaNK}) one
should then sum individual contributions from all these wave vectors: 

\begin{equation}
P=\sum_{n}\sum_{\nu}\int_{0}^{4\pi}d\Omega\left(\frac{\pi^{2}\omega_{0}^{2}}{V}\frac{\left|\mathbf{A^{\nu}}_{n\mathbf{k}}(\mathbf{r}_{0})\cdot\mathbf{d}\right|^{2}}{J_{n\mathbf{k}}^{\nu}\left|\mathbf{V}_{n\mathbf{k}}^{\nu}\right|}\frac{\left|\mathbf{k}_{n}^{\nu}\right|^{2}}{\cos\varphi}\right),\label{math:powerOmega1}\end{equation}

\begin{figure}[tb]
\begin{center}\includegraphics[%
  width=0.95\columnwidth]{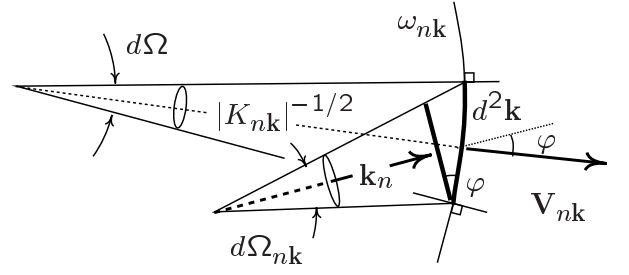}\end{center}

\caption{Diagram to derive the relation between solid angles in the \textbf{k}-space
and coordinate space. The iso-frequency contour for frequency $\omega_{n\mathbf{k}}$
is presented. The Jacobian of the transformation (\ref{math:mappingTheta}-\ref{math:mappingPhi})
is given by the ratio $d\Omega/d\Omega_{n\mathbf{k}}$. By the definition
of the solid angle, the solid angle in \textbf{k}-space is $d\Omega_{n\mathbf{k}}=d^{2}\mathbf{k}\cos\varphi/\left|\mathbf{k}_{n}\right|^{2}$,
while the corresponding solid angle in coordinate space is $d\Omega=d^{2}\mathbf{k}\left|K_{n\mathbf{k}}\right|$.
That gives the Jacobian $J_{n\mathbf{k}}=\left|\mathbf{k}_{n}\right|^{2}\left|K_{n\mathbf{k}}\right|/\cos\varphi$.
Here $\varphi$ is an angle between the wave vector and the group
velocity vector. $d^{2}\mathbf{k}$ is the surface element of the
iso-frequency surface.\label{fig:jacobian}}
\end{figure}

The geometrical relationship between solid angles in \textbf{k}-space
and coordinate space (Fig.~\ref{fig:jacobian}) results in the following
formula for the Jacobian \textbf{}(\ref{math:jacobian})\textbf{\[
J_{n\mathbf{k}}^{\nu}=\left|\mathbf{k}_{n}^{\nu}\right|^{2}\left|K_{n\mathbf{k}}^{\nu}\right|/\cos\varphi.\]
}Then, equation (\ref{math:powerOmega1}) can be transformed to the
form:\begin{equation}
P=\int_{0}^{4\pi}d\Omega\left(\sum_{n}\sum_{\nu}\frac{\pi^{2}\omega_{0}^{2}}{V}\frac{\left|\mathbf{A}_{n\mathbf{k}}^{\nu}(\mathbf{r}_{0})\cdot\mathbf{d}\right|^{2}}{\left|\mathbf{V}_{n\mathbf{k}}^{\nu}\right|\left|K_{n\mathbf{k}}^{\nu}\right|}\right),\label{math:totalPowerFinal}\end{equation}
where $\mathbf{V}_{n\mathbf{k}}^{\nu}=\nabla_{\mathbf{k}}\omega_{n\mathbf{k}}$
is the group velocity and $K_{n\mathbf{k}}^{\nu}$ determines the
Gaussian curvature of the iso-frequency surface at the stationary
point $\mathbf{k}_{n}=\mathbf{k}_{n}^{\nu}$. Finally, the radiated
power of the dipole per solid angle in coordinate space is given by
the function enclosed in the brackets in (\ref{math:totalPowerFinal})\begin{equation}
\frac{dP}{d\Omega}=\sum_{n}\sum_{\nu}\frac{\pi^{2}\omega_{0}^{2}}{V}\frac{\left|\mathbf{A}_{n\mathbf{k}}^{\nu}(\mathbf{r}_{0})\cdot\mathbf{d}\right|^{2}}{\left|\mathbf{V}_{n\mathbf{k}}^{\nu}\right|\left|K_{n\mathbf{k}}^{\nu}\right|}.\label{math:angularFinal}\end{equation}

Formula (\ref{math:angularFinal}) provides a simple route to calculate
an angular distribution of radiated power of the point dipole (\ref{math:sourceTerm})
inside a photonic crystal. It can be interpreted as a decay rate,
at which the dipole transfers energy to the electromagnetic waves
with the group velocity in the observation direction. Then, $(d\Gamma/d\Omega)=(dP/d\Omega)/\hbar\omega_{0}$
is related to the probability of the radiative transition of an excited
atom with emitting a photon traveling in the given observation direction. 

Basically, formulae (\ref{math:totalPowerFinal}-\ref{math:angularFinal})
involve calculations of the Bloch wave vectors $\mathbf{k}_{n}^{\nu}$,
ending at the iso-frequency surface $\omega_{n\mathbf{k}}=\omega_{0}$,
the corresponding group velocity vectors $\mathbf{V}_{n\mathbf{k}}^{\nu}$,
the Gaussian curvature of the iso-frequency surface $K_{n\mathbf{k}}^{\nu}$
and the local coupling strength of the dipole moment with a Bloch
eigenwave $(n,\mathbf{k})$, given by the factor $\left|\mathbf{A}_{n\mathbf{k}}^{\nu}(\mathbf{r}_{0})\cdot\mathbf{d}\right|$.
The primary difficulty in obtaining the coordinate space distribution
of radiated power $(dP/d\Omega)$ (\ref{math:angularFinal}) is that
the wave vector, the group velocity and the Gaussian curvature are
all functions of the \textbf{k}-space direction. Whereas an angular
dependence of the radiative power $(dP/d\Omega)$ is given by the
corresponding group velocity direction $(\theta,\phi)$. To calculate
the radiated power $(dP/d\Omega)$ (\ref{math:angularFinal}) one
should take an inverse of the mapping (\ref{math:mappingTheta}-\ref{math:mappingPhi}).
This inverse is not necessarily unique. In the case of multiple stationary
points (\ref{math:stationaryOmega}-\ref{math:stationaryK}), one
direction $(\theta,\phi)$ results from several different \textbf{k}-space
directions $(\theta_{\mathbf{k}},\phi_{\mathbf{k}})$ (Fig.~\ref{fig:folding}).
This requires that the inversion of the mapping (\ref{math:mappingTheta}-\ref{math:mappingPhi})
must be done point-by-point.

As a simple exercise, formula (\ref{math:angularFinal}) is applied
here to calculate an angular distribution of power radiated by a dipole
in free space. The wave vector and the group velocity in free space
are parallel and their values are simply given by $\left|\mathbf{k}\right|=\omega_{0}/c$
and $c$, respectively. The Gaussian curvature of the iso-frequency
surface is a square of the inverse wave vector $1/\left|\mathbf{k}\right|^{2}$.
And the appropriate normal modes are plane waves \[
\mathbf{A}_{n\mathbf{k}}(\mathbf{r})=\sqrt{\frac{V}{(2\pi)^{3}}}e^{i\mathbf{k\cdot r}}\widehat{\mathbf{a}}_{n\mathbf{k}},\]
where $\widehat{\mathbf{a}}_{n\mathbf{k}}$ is a polarization vector
orthogonal to the wave vector $\mathbf{k}$. Then, the radiated power
is given by (\ref{math:angularFinal})\begin{eqnarray}
\left(\frac{dP}{d\Omega}\right)_{free} & = & \frac{1}{8\pi}\frac{\omega_{0}^{4}}{c^{3}}\left|\mathbf{d}\right|^{2}\sin^{2}\theta\label{math:PowerFreeSpace}\end{eqnarray}
 yielding the usual results for radiation pattern in free space \cite{bookJackson}.

\section{\label{sec:photon_focusing}Photon focusing}

The factor $\left|\mathbf{A}_{n\mathbf{k}}^{\nu}(\mathbf{r}_{0})\cdot\mathbf{d}\right|^{2}$
in relation (\ref{math:angularFinal}), giving the coupling strength
of dipole moment with the photonic crystal eigenmode at the dipole
position, can display a complex angular behavior, which depends on
eigenmode structure and dipole orientation with respect to the crystal
lattice. To study the net result of the influence of the photonic
crystal on the radiation pattern of the emitter, it is instructive
to model an isotropic light source producing a uniform distribution
of wave vectors. Moreover, an isotropic point source is usually a
good model for a common experimental situation of emitters with random
distribution of dipole moment (dye molecules \cite{gaponenko99,schiemer00,romanov01},
quantum dots \cite{gaponenko99,romanov03}, etc.). Then, the radiated
power (\ref{math:angularFinal}) should be averaged over the dipole
moment orientation, which simply yields a factor of $\left|\mathbf{d}\right|^{2}/3$\begin{equation}
\left(\frac{dP}{d\Omega}\right)_{i}=\sum_{n}\sum_{\nu}\frac{\left(2\pi c\right)^{3}}{V\omega_{0}^{2}}\frac{\left|\mathbf{A}_{n\mathbf{k}}^{\nu}(\mathbf{r}_{0})\right|^{2}}{\left|\mathbf{V}_{n\mathbf{k}}^{\nu}\right|\left|K_{n\mathbf{k}}^{\nu}\right|}.\label{math:isotropicAngular0}\end{equation}
Here the result was normalized to the radiated power in free space.
Now, the factor $\left|\mathbf{A}_{n\mathbf{k}}^{\nu}(\mathbf{r}_{0})\right|^{2}$
gives a field strength at the source position and has no angular dependence.
So, the radiation pattern of a point isotropic emitter is defined
by\begin{equation}
\left(\frac{dP}{d\Omega}\right)_{i}\sim\sum_{n}\sum_{\nu}\left|\mathbf{V}_{n\mathbf{k}}^{\nu}\right|^{-1}\left|K_{n\mathbf{k}}^{\nu}\right|^{-1}.\label{math:isotropicAngular}\end{equation}

The radiated power (\ref{math:isotropicAngular}) is proportional
to the inverse group velocity, $\left|\mathbf{V}_{n\mathbf{k}}^{\nu}\right|^{-1}$,
and to the inverse Gaussian curvature, $\left|K_{n\mathbf{k}}^{\nu}\right|^{-1}$
of iso-frequency surface.
A large enhancement of emission rate is expected when the group velocity
is small. This can be interpreted as a consequence of the long interaction
time of the emitter and the radiation field \cite{dowling94,nojima98,sakoda99}.
In a similar fashion, a small Gaussian curvature formally implies
an enhancement of radiated power along a certain observation direction.
While spontaneous emission enhancement due to a small group velocity
involves non-linear interaction of radiation and emitter, the enhancement
due to a small Gaussian curvature is a linear phenomenon related to
the anisotropy of the photonic crystal and is a result of the beam
steering effect. Being a measure of the rate, with which emitter transfers
energy in photons with a given group velocity, radiated power (\ref{math:isotropicAngular})
will be enhanced if many photons with different wave vectors reach
the same detector. The enhancement of the radiated power, which is
due to the small Gaussian curvature is called \emph{photon focusing}
\cite{etchegoin96,chigrin2001} and has a major influence on radiation
pattern of a point source in a photonic crystal. 

The physical picture of the \emph{photon focusing} can be illustrated
in the following manner (Fig.~\ref{fig:folding}). An iso-frequency
surface of an isotropic and homogeneous medium is a sphere. There
is only one stationary point with $\widehat{\mathbf{x}}\cdot\mathbf{V}_{n\mathbf{k}}^{\nu}>0$
and thus only one wave propagating in the given direction. Figure
\ref{fig:folding}-left is an example of a part of the actual iso-frequency
contour of a 2D photonic crystal made out of dielectric rods placed
in vacuum (see Sec.~\ref{sec:example_2D_PhC} for further details).
The anisotropy of the crystal implies a complex non-spherical iso-frequency
surface, which can have several stationary points with $\widehat{\mathbf{x}}\cdot\mathbf{V}_{n\mathbf{k}}^{\nu}>0$
(Fig.~\ref{fig:folding}-left). Several waves can propagate in a
given direction inside a photonic crystal. It is illustrative to construct
the \emph{wave surface} in coordinate space. To construct the wave
surface one should plot a ray in the observation direction $\widehat{\mathbf{x}}$
starting from the point source position and having the length of the
group velocity $\left|\mathbf{V}_{n\mathbf{k}}^{\nu}\right|$. An
example of the wave contour is presented in figure \ref{fig:folding}-right.
The existence of multiple stationary point implies that the wave surface
is a complex multivalued surface parameterized by wave vector $\mathbf{k}_{n}$.
Figure \ref{fig:folding} illustrates how this can result in a fold
of the wave surface.

In the vicinity of the parabolic point with zero Gaussian curvature
an iso-frequency surface is flat. That implies, that a very large
number of eigenwaves with wave vectors in the vicinity of a parabolic
point have nearly the same group velocity, contributing to the energy
flux in the direction parallel to that group velocity. In the figure
\ref{fig:folding}, it is illustrated by mapping two equal solid angle
sections along different observation direction in the coordinate space
onto the corresponding solid angle sections in \textbf{k}-space \cite{wolfe80}.
The black solid angle section in coordinate space maps onto a single
smaller solid angle section in \textbf{k}-space implying a {}``defocusing''
of the energy flux. The shaded solid angle section in coordinate space,
which crosses three different branches of the wave contour, maps onto
two different and larger solid angle sections in \textbf{k}-space
implying enhancement ({}``focusing'') of the energy flux in this
group velocity direction. This results in strongly varying angular
distribution of emission intensity with sharp singularities (caustics).

\section{\label{sec:example_2D_PhC}Numerical example: 2D photonic crystal}

In this section the theoretical approach developed in the previous
sections is applied to the numerical calculation of the radiation
pattern of a point source placed inside a 2D photonic crystal. A point
source is situated inside the crystal and it produces an isotropic
and uniform distribution of wave vectors $\mathbf{k}_{n}$ with the
frequency $\omega_{0}$. 

\begin{figure}[tb]
\begin{center}\includegraphics[%
  width=0.95\columnwidth]{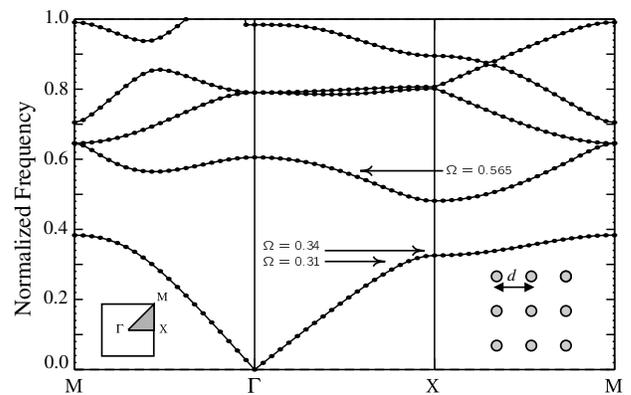}\end{center}

\caption{Photonic band structure of TM modes for the square lattice photonic
crystal with refractive index of the rods $n=2.9$, the lattice constant
d  and radius of the rods 0.15d. The frequency
is normalized to $\Omega=\omega d/2\pi c=d/\lambda$. $c$ is the
speed of light in vacuum. Insets show the first Brillouin zone of
the crystal with the irreducible zone shaded light gray (left) and
a part of the lattice (right).\label{fig:bandStructure}}
\end{figure}
\begin{figure}[tb]
\begin{center}\includegraphics[%
  width=0.80\columnwidth,
  keepaspectratio]{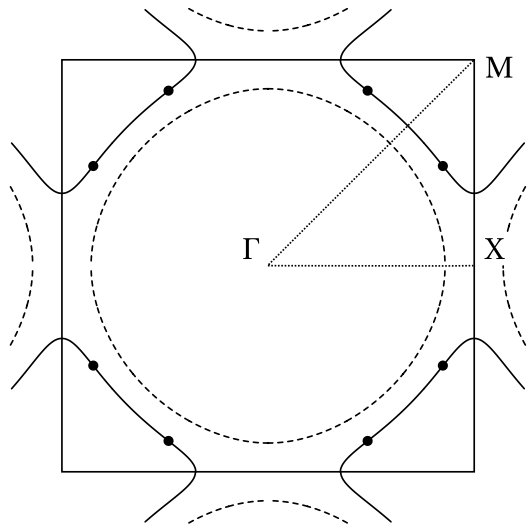}\end{center}

\caption{Iso-frequency contours of the square lattice photonic crystal for
the normalized frequencies $\Omega=0.31$ (dashed line) and $\Omega=0.34$
(solid line). Parabolic point are marked by the black dots. The first
Brillouin zone of the lattice is plotted in order to show the spatial
relation between zone boundary and iso-frequency contours.\label{fig:IFS-B1}}
\end{figure}

An infinite 2D square lattice of dielectric rods in vacuum (Fig.~\ref{fig:bandStructure})
is considered in the case of in-plane propagation. Consequently, the
problem of an electromagnetic wave interaction with a 2D photonic
crystal is reduced to two independent problems, which are called TE
and TM, when the electric or magnetic field is parallel to the axis
of the rods. In the illustrative example presented in this section,
all numerical calculations have been performed for TM modes of the
crystal. The photonic band structure of the crystal made of the rods
with the refractive index $n=2.9$ is presented in the figure~\ref{fig:bandStructure}.
The band structure has been calculated using the plane wave expansion
method \cite{johnson2001}.

\begin{figure}[tb]
\begin{center}\includegraphics[%
  width=0.80\columnwidth,
  keepaspectratio]{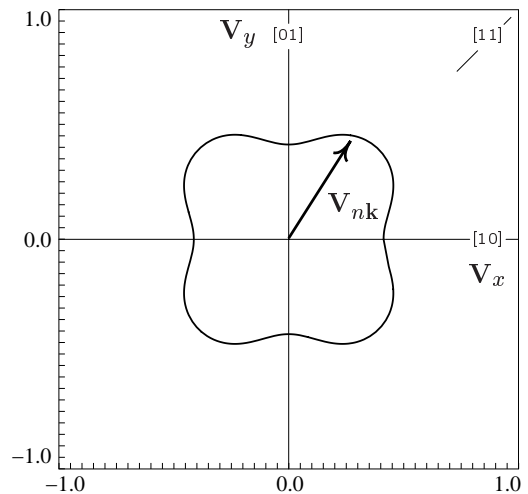}\end{center}

\caption{Wave contour corresponding to the normalized frequency $\Omega=0.31$.
The group velocity is plotted in the units of the speed of light in
vacuum. High symmetry directions of the square lattice are specified.\label{fig:WC-0.31}}
\end{figure}

In the figure~\ref{fig:IFS-B1} iso-frequency contours of the crystal
are presented for two frequencies belonging to the first photonic
band (Fig.~\ref{fig:bandStructure}). To plot an iso-frequency contour,
the photonic band structure for all wave vectors within the irreducible
BZ was calculated and then the equation $\omega(\mathbf{k})=\omega_{0}$
was solved for a given frequency $\omega_{0}$. Frequencies have been
chosen below ($\Omega=0.31$) and above ($\Omega=0.34$) the low edge
frequency of the stopband in the $\Gamma\mathrm{X}$ direction of
the crystal. The iso-frequency contours below and above the stopband
edge frequency show an important difference. As the frequency stays
below the stopband, an iso-frequency contour is \emph{closed} and
almost circular (Fig.~\ref{fig:IFS-B1}). The corresponding wave
contour (see Section \ref{sec:photon_focusing} for definition) is
presented in the figure \ref{fig:WC-0.31}. To calculate the group
velocity, the plane wave expansion method \cite{johnson2001} and
the Hellmann-Feyman theorem were used. The group velocity $\left|\mathbf{V}_{n\mathbf{k}}^{\nu}\right|$
and the Gaussian curvature $\left|K_{n\mathbf{k}}^{\nu}\right|$ of
the iso-frequency contours are relatively slow function of the wave
vector. The Gaussian curvature does not vanish for any wave vector.
This implies a small anisotropy in the energy flux inside the crystal.

\begin{figure}[tb]
\begin{center}\includegraphics[%
  width=0.80\columnwidth]{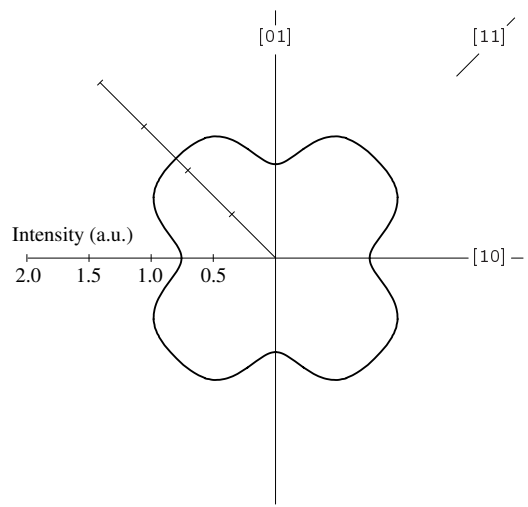}\end{center}

\caption{Angular distribution of radiative power corresponding to the normalized
frequency $\Omega=0.31$. High symmetry directions of the square lattice
are specified. \label{fig:RP-0.31}}
\end{figure}

To find how a radiated power varies in coordinate space, one should
calculate the group velocity and the Gaussian curvature on the iso-frequency
contour $\omega(\mathbf{k})=\omega_{0}$ as functions of an angle
in coordinate space. As the wave contour is single valued function,
the inverse of the mapping (\ref{math:mappingTheta}-\ref{math:mappingPhi})
from \textbf{k}-space to coordinate space is one-to-one and can be
easily done. In the figure \ref{fig:RP-0.31} the polar plot of radiated
power is presented, which shows small amount of anisotropy. The angular
distribution of the radiated power possesses four-fold rotational
symmetry of the crystal.

\begin{figure}[tb]
\begin{center}\includegraphics[%
  width=0.80\columnwidth]{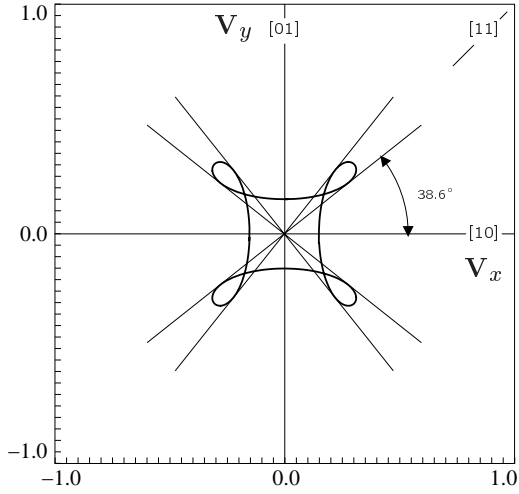}\end{center}

\caption{Wave contour corresponding to the normalized frequency $\Omega=0.34$.
The group velocity is plotted in the units of the speed of light in
vacuum. The directions corresponding to the folds of the wave contour
are shown.\label{fig:WC-0.34}}
\end{figure}

With increase of the frequency up to the stopband, the topology of
the iso-frequency contour abruptly changes. The stopband developed
in the $\Gamma\mathrm{X}$ direction and the iso-frequency contour
becomes \emph{open} (Fig.~\ref{fig:IFS-B1}). This topology changes
result in complex contour with alternating regions of different Gaussian
curvature sign. Parabolic points, where the Gaussian curvature vanishes,
are marked by black dots in the figure \ref{fig:IFS-B1}. As it has
been discussed in section \ref{sec:photon_focusing}, vanishing curvature
results in the folds of the wave contour. The wave contour corresponding
to the iso-frequency $\Omega=0.34$ is presented in the figure \ref{fig:WC-0.34}.
A pair of the parabolic points in the first quarter of the Brillouin
zone results in a cuspidal structure of the wave contours in the first
quarter of the coordinate space. This dramatically increases anisotropy
of the energy flux.

\begin{figure}[tb]
\begin{center}\includegraphics[%
  width=0.80\columnwidth]{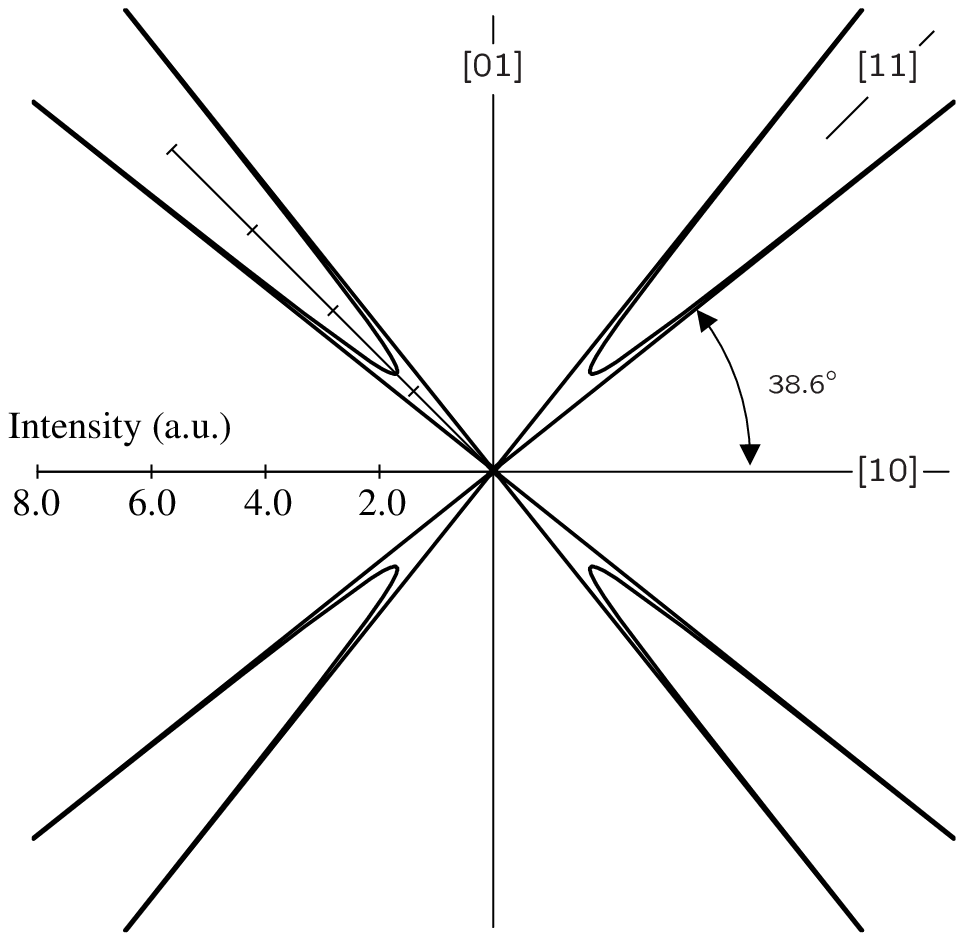}\end{center}

\caption{Angular distribution of radiative power corresponding to the normalized
frequency $\Omega=0.34$. The directions of infinite radiative power
(caustic) coincide with the directions of the folds of the wave contour
(Fig.~\ref{fig:WC-0.34}).\label{fig:RP-0.34}}
\end{figure}

\begin{figure}[tb]
\begin{center}\includegraphics[%
  width=0.80\columnwidth,
  keepaspectratio]{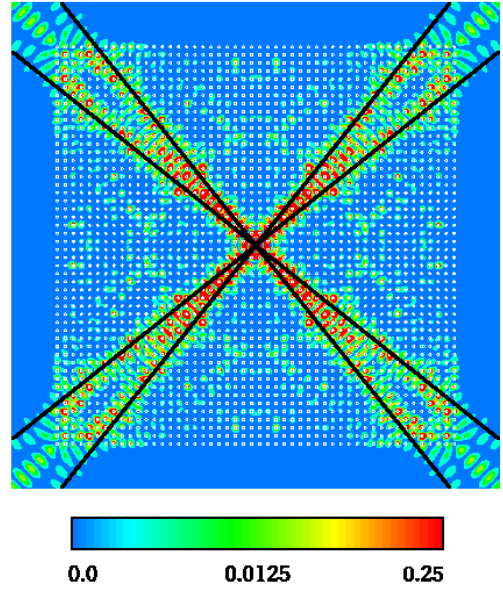}\end{center}

\caption{FDTD calculation. Map of the modulus of the Poynting vector field
for a $50\times50$ rod photonic crystal excited by a point isotropic
source with the normalized frequency $\Omega=0.34$. The location
of the crystal in the simulation is shown together with asymptotic
directions of photon focusing caustics.\label{fig:fdtd}}
\end{figure}

The folds in the wave contours yields that inverse of the mapping
(\ref{math:mappingTheta}-\ref{math:mappingPhi}) from \textbf{k}-space
to coordinate space is not one-to-one any more. To apply the formula
(\ref{math:isotropicAngular}) to calculate an angular distribution
of radiated power in such a case, one should proceed as follows. At
first, the Gaussian curvature as a function of the wave vector should
be calculated. Then, wave vectors and group velocities corresponding
to the parabolic points on the iso-frequency surface should be found.
An inversion of the mapping (\ref{math:mappingTheta}-\ref{math:mappingPhi})
should be calculated separately for each of the branches of the wave
contour. The total radiated power is a sum of the different contributions
from these branches. In the figure \ref{fig:RP-0.34} the polar plot
of radiated power (\ref{math:isotropicAngular}) corresponding to
the normalized frequency $\Omega=0.34$ is presented. The energy flux
is strongly anisotropic for this frequency, showing relatively small
intensity in the directions of the stopband, and infinite intensity
(caustics) in the directions of the folds.

To substantiate this behavior, finite difference time domain (FDTD)
calculations were done \cite{taflove95,sullivan02}. The simulated
structure was a $50\times50$ lattice of dielectric rods in vacuum.
The crystal is surrounded by an extra $5d$ wide layer of a free space.
The simulation domain was discretized into squares with a side $\Delta=d/32$.
The total simulation region was $1920\times1920$ cells plus 8-cell
wide perfectly matched layer (PML) \cite{berenger94}. The point isotropic
light source was modeled by a soft source \cite{taflove95,sullivan02}
with a homogeneous spacial dependence and sinusoidal temporal dependence
of the signal. All FDTD calculations was performed with a commercial tool \cite{ise}.

In figure~\ref{fig:fdtd} the map of the modulus of the Poynting
vector field is shown. The point source is placed in the middle of
the crystal. The field map is shown for one instant time step. The
snap-shots were captured after 10000 time steps, where the time step
was $4.38\times10^{-17}$ s (0.99 of the Courant value). The structure
of the crystal is superimposed on the field map. From figure~\ref{fig:fdtd}
one can see, that the emitted light is focused in the directions coinciding
with the predicted directions of the folds (black lines). 

\begin{figure}[tb]
\begin{center}\includegraphics[%
  width=0.80\columnwidth]{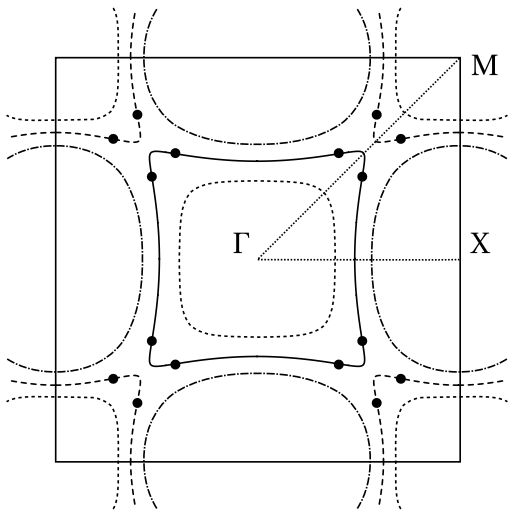}\end{center}

\caption{Iso-frequency contours of the square lattice photonic crystal for
the normalized frequencies $\Omega=0.55$ (dotted line), $\Omega=0.565$
(solid and dashed line) and $\Omega=0.58$ (dashed-dotted line). Two
branches of the iso-frequency contour of $\Omega=0.565$ is plotted
as solid and dashed lines. Parabolic point are marked by the black
dots. The first Brillouin zone of the lattice is plotted in order
to show the spatial relation between zone boundary and iso-frequency
contours.\label{fig:IFS-B2}}
\end{figure}

\begin{figure}[tb]
\begin{center}\includegraphics[%
  width=0.80\columnwidth]{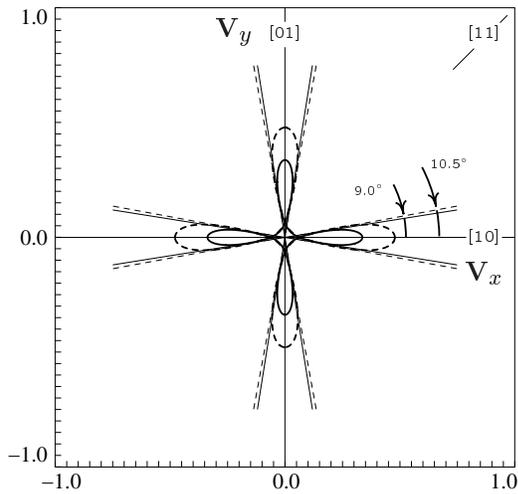}\end{center}

\caption{Wave contours corresponding to the normalized frequency $\Omega=0.565$.
Solid (dashed) wave contour corresponds to solid (dashed) iso-frequency
contour in figure \ref{fig:IFS-B2}. The group velocity is plotted
in the units of the speed of light in vacuum. The directions corresponding
to the folds of the wave contour are shown.\label{fig:WC-0.565}}
\end{figure}

\begin{figure}[tb]
\begin{center}\includegraphics[%
  width=0.80\columnwidth]{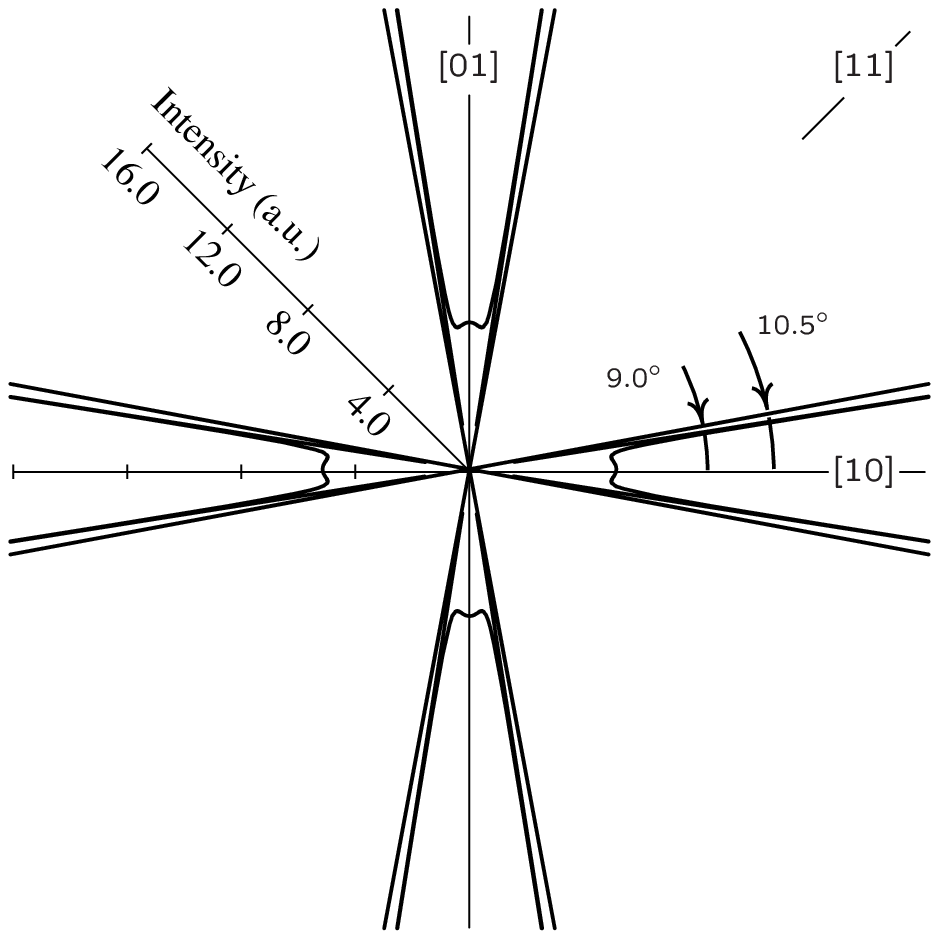}\end{center}

\caption{Angular distribution of radiative power corresponding to the normalized
frequency $\Omega=0.565$. The directions of infinite radiative power
(caustic) coincide with the directions of the folds of the wave contour.
\label{fig:RP-0.565}}
\end{figure}

In figures~\ref{fig:IFS-B2}-\ref{fig:RP-0.565}, a more complicated
example of the anisotropy of a photonic crystal is presented. Iso-frequency
contours for three frequencies belonging to the second photonic band
of the crystal are plotted in the figure~\ref{fig:IFS-B2}. While
iso-frequency contours for the normalized frequencies $\Omega=0.55$
and $\Omega=0.58$ have non-vanishing Gaussian curvature for all wave
vectors leading to only limited anisotropy of the energy flux, the
iso-frequency contour for the normalized frequencies $\Omega=0.565$
displays several parabolic points. Moreover, the iso-frequency contour
consists of two branched with slightly different shapes (solid and
dashed lines in the figure~\ref{fig:IFS-B2}). Two branches yield
two wave contours with cuspidal folds in coordinate space (Fig.~\ref{fig:WC-0.565}).
Applying the formula (\ref{math:isotropicAngular}) to the radiative
power calculation, one should sum over contributions coming from all
branches of the wave contours in coordinate space. An angular distribution
of radiative power for the normalized frequencies $\Omega=0.565$
is presented in the figure~\ref{fig:RP-0.565}. Within the first
quarter of the coordinate space, four caustics with infinite radiated
power present in the energy flux corresponded to four parabolic points
on two branches of the iso-frequency contours.

\section{\label{sec:dipole:summary}Summary}

In this paper, by analyzing a dipole field in the radiation zone it
was shown, that the principal contribution to the far-field of the
dipole radiating in a photonic crystal comes from the regions of the
iso-frequency surface in the wave vector space, at which the eigenwave
group velocity is parallel to observation direction $\mathbf{\widehat{x}}$.
It was also shown that anisotropy of a photonic crystal reveals itself
in the strongly non-spherical wave front leading to modifications
of both far-field radiation pattern and spontaneous emission rate.
By systematic analysis of the Maxwell equations a simple formula to
calculate an angular distribution of radiated power due to a point
dipole placed in a photonic crystal was derived. The formula only
involves calculations of the wave vectors, the group velocity, the
coupling strength of the dipole moment with the field and the Gaussian
curvature on the iso-frequency surface corresponding to the frequency
of the oscillating dipole. That can be done by simple plane wave expansion method
and is not computationally demanding. A numerical example was given
for a square-lattice 2D photonic crystal. It was shown by applying
developed formalism and substantiated by FDTD calculations, that if
a dipole frequency is within a partial photonic bandgap, a far-field
radiation pattern is strongly modified with respect to the dipole
radiation pattern in vacuum, demonstrating suppression in the directions
of the spatial stopband and enhancement in the direction of the group
velocity, which is stationary with respect to a small variation of
the wave vector.

\section{Acknowledgments}

This work was partially supported by the EU-IST project APPTech IST-2000-29321
and German BMBF project PCOC 01 BK 253.

\end{document}